\documentclass[onecolumn,preprintnumbers,amsmath,amssymb]{revtex4}

\begin{document}


\title{WKB formalism and a lower limit for the energy eigenstates of bound states for some potentials}
\author{Luis F. Barrag\'{a}n-Gil}
\email{brgl@xanum.uam.mx} \affiliation{Departamento de
F\'{\i}sica,
Universidad Aut\'onoma Metropolitana--Iztapalapa\\
Apartado Postal 55--534, C.P. 09340, M\'exico, D.F., M\'exico.}

\author{Abel Camacho}
\email{acq@xanum.uam.mx} \affiliation{Departamento de F\'{\i}sica,
Universidad Aut\'onoma Metropolitana--Iztapalapa\\
Apartado Postal 55--534, C.P. 09340, M\'exico, D.F., M\'exico.}

\date{\today}

\begin{abstract}
In the present work the conditions appearing in the so--called WKB
approximation formalism of quantum mechanics are analyzed. It is
shown that, in general, a careful definition of an approximation
method requires the introduction of two length parameters, one of
them always considered in the text books on quantum mechanics,
whereas the second one is usually neglected. Afterwards we define
a particular family of potentials and prove, resorting to the
aforementioned length para\-meters, that we may find an energy
which is a lower bound to the ground energy of the system. The
idea is applied to the case of a harmonic oscillator and also to a
particle freely falling in a homogeneous gravitational field, and
in both cases the consistency of our method is corroborated. This
approach, together with the so--called Rayleigh--Ritz formalism,
allows us to define an energy interval in which the ground energy
of any potential, belonging to our family, must lie.

\end{abstract}

\maketitle

\section{Introduction}

In quantum mechanics the situation is, in a certain sense, quite
similar to the situation in classical mechanics, namely, the
number of physical systems of interest for which the corresponding
motion equations can be solved exactly is not very large.
Therefore approximation methods have been developed and play an
important role in quantum mechanics. Among these approaches we may
mention the WKB, Rayleigh--Ritz, or perturbation methods
\cite{[1]}. These different ideas have realms of applicability
that not always intersect. For instance, WKB is applicable to
states of the corresponding system characterized by very large
values of a certain quantum number \cite{[2]}, and the variational
method, also known as Rayleigh--Ritz, is employed to find bounds
to the ground state energy of a quantum system \cite{[2]}.

In the present paper we will focus on the WKB method and analyze
the conditions that appear in the implementation of it. It will be
shown that, in general, they involve two conditions. Usually only
one is considered, the remaining one is somehow neglected. Indeed,
in all text books on quantum mechanics the fact that the method
yields a good approximation only for the case in which there are
several wavelengths between the corresponding classical turning
points is very carefully explained, and mathematically implemented
\cite{[2]}.

Nevertheless, there is an additional approximation that is always
done, and implicitly introduced, namely, at a point on the
potential curve a Taylor expansion for the potential is done, and
then only a finite number of terms of this expansion are
considered (usually two terms \cite{[1], [2]}). Then this
approximate Schr\"odinger equation is solved, and with it an
auxiliary function is defined. This auxiliary function is employed
to find the relation between the coefficients of the two involved
WKB functions, one function on the left--hand side and a second
one on the right--hand side of the classical turning point.

It has to be underlined that when this is done, in order to
implement the method the solution (stemming from the approximate
Schr\"odinger equation) is compared against the WKB wavefunction.
This procedure is carried out on both sides of the corresponding
classical turning point. We must distinguish, in connection with
the approximate Schr\"odinger equation and its corresponding
solution, that the validity region of the approximate equation
(remember that it is obtained truncating the Taylor expansion for
the potential) does define the region in which the corresponding
solution can be applied. Of course, the solution to the
approximate Schr\"odinger equation may have a definition domain
larger than the region in which the approximate potential is
valid, but if we use the solution at points outside the validity
region of the approximate Schr\"odinger equation, then we use the
solution to a potential that does not represent very good the
physical situation. In other words, a rigorous procedure implies
that the solution to the approximate Schr\"odinger equation has to
be compared against the WKB wavefunction within the validity
region of the aforementioned differential equation.

This last remark implies that there are two different length
parameters involved in the WKB procedure: (i) the first one tells
us that we cannot be very close to the classical turning point,
otherwise the approximate wavefunction diverges \cite{[1], [2]};
(ii) the second length parameter appears when we recognize that
the comparison between the solution to the approximate
Schr\"odinger equation and the WKB wavefunction has to be carried
out within the validity region of the approximate Schr\"odinger
equation. Usually, only the first length parameter is considered
in the text books on quantum mechanics \cite{[1], [2]}, the second
one is always neglected, and we may wonder if a careful analysis
of it is necessary, since we seek an approximate solution, valid
at both sides of our classical turning points.

In the present work we will take into account both conditions and
show that a careful analysis of them leads us to some interesting
facts. For instance, we will show that it is possible to find a
lower bound to the ground energy of the so--called bound states
for some potentials. This is an interesting result, since the
current approximation methods cannot render a lower bound to the
ground state, only the Rayleigh--Ritz method provides a formalism
which allows us to obtain, via an energy functional, an energy
which cannot be smaller than ground energy of the corresponding
system. With our method the ground energy is always larger than
the energy that our method provides. Clearly, joining the
conclusions of Rayleigh--Ritz with ours we may find an interval in
which the ground energy of a one--dimensional system has to lie.
Though the present approach is restricted to a certain family of
potentials we will prove that the case of a harmonic oscillator,
or a particle freely falling in a homogeneous gravitational field,
among other potentials, belong to this family.

\section{Validity region for the WKB formalism}

The core point in the WKB formalism is related to the fact that
the coefficients appearing in the different regions, which are
separated by the corresponding classical turning points, are
matched  resorting to the so--called connection formulas
\cite{[1]}. These formulas are obtained introducing an approximate
Schr\"odinger equation at the classical turning point, solving it,
finding its asymptotic behavior, and comparing the coefficients
against those stemming from the WKB formalism. The phrase {\it an
approximate Schr\"odinger equation at the classical turning point}
plays here a very relevant role, and has to be explained in a very
clear manner.

Let us now proceed to do this. Consider now one of the classical
turning points, say $x_0$. A Taylor expansion at this point, for
the potential is introduced ($V: I\rightarrow \Re$, where
$I\subseteq\Re$)

\begin{equation}
V(x) = V(x_0) + V^{'}(x_0)(x-x_0) +
\frac{V^{''}(x_0)}{2!}(x-x_0)^2+.... \label{Approxpotential1}
\end{equation}

The approximate time--independent Schr\"odinger equation is
defined resorting to the first non--vanishing derivative at $x_0$,
usually the first--order derivative

\begin{equation}
\frac{-\hbar^2}{2m}\frac{d^2\psi}{dx^2} + \Bigl[V(x_0) +
V^{'}(x_0)(x-x_0)\Bigr]\psi(x) = E\psi(x).
\label{Approxdiffeqaution1}
\end{equation}

The solutions to (\ref{Approxdiffeqaution1}) are Airy functions
\cite{[1]}. The second step is to take the asymptotic form for the
corresponding solutions to (\ref{Approxdiffeqaution1}). This has
to be done because the main idea is to use this approximate
equation as a patching expression, and employ it to connect the
expressions stemming from the WKB formalism on both sides of each
one of the classical turning points. Here the verb {\it connect}
means to find the relations between the coefficients of the
solutions obtained with WKB on both sides of the classical turning
point. Since this part of the method requires the WKB
wavefunctions, then the comparison between the Airy functions and
the WKB wavefunctions has to be done in a region where the latter
are well defined, i.e., sufficiently far from the classical
turning point \cite{[1]}.

Up to this point everything is quite clear in the textbooks on
quantum mechanics. Nevertheless, there is an issue which is not
addressed, and is a relevant one. Namely, the comparison has to be
carried out far enough from the classical turning point, but
resorting to the asymptotic behavior of the Airy functions this
last argument is not enough. Indeed, we must be sure that in the
region where the comparison is done the approximation chosen for
the Schr\"odinger equation is a good one, otherwise we would be
employing the asymptotic behavior of the solutions of an equation
which is not a good approximation in the region where the
comparison is done. In other words, a careful procedure should
check that the region in which WKB wavefunctions are valid
contains as a subset an interval in which
(\ref{Approxdiffeqaution1}) is valid, i.e., we do not need to take
into account the second order derivative, for instance.

This last statement can be rephrased asserting that the region of
the comparison is defined as follows (here from the very beginning
we assume the usual case, a linear approximation for the
Schr\"odinger equation, additionally recall that at the classical
turning point $E=V(x_0)$)

\begin{equation}
V(x) \approx E + V^{'}(x_0)(x-x_0). \label{Approxpotential2}
\end{equation}

If this last approximation is a good one, then the terms of the
form $(x-x_0)^k$, with $k\geq 2$, can be neglected and this
assertion implies

\begin{equation}
\vert V^{'}(x_0)(x-x_0)\vert>\vert
\frac{V^{''}(x_0)}{2!}(x-x_0)^2\vert. \label{Conditionpotential1}
\end{equation}

This last expression defines the validity region of
(\ref{Approxdiffeqaution1}). Indeed, (\ref{Conditionpotential1})
entails that

\begin{equation}
\vert 2\frac{V^{'}(x_0)}{V^{''}(x_0)}\vert>\vert(x-x_0)\vert.
\label{Interval1}
\end{equation}

In other words, if the matching is done in a region which violates
this last condition, then the approximate Schr\"odinger equation
should consider higher--order derivatives. Notice that the
condition is a function of the involved classical turning point,
$x_0$, in other words, the size of the validity region depends
upon $x_0$, it is not, in the general case, constant.

The use of the WKB method requires an additional condition
\cite{[1]}, the one can be tracked down to the fact that the
wavelength of the corresponding particle has to be much smaller
than the region in which the potential has a noticeable change in
its value

\begin{equation}
\frac{2\vert E- V(x)\vert}{\vert dV/dx\vert} >> \lambdabar =
\frac{\hbar}{\sqrt{2m[E - V(x)]}}. \label{WKB1}
\end{equation}

Harking back to (\ref{Approxpotential1}) we impose the condition
that

\begin{equation}
limsup\vert\frac{ V^{(l+1)}(x_0)}{V^{(l)}(x_0)}(x-x_0)\vert <1.
\label{Convergence1}
\end{equation}

This is tantamount to ask for the absolute convergence of the
Taylor series of $V(x)$ \cite{[3]}. Since this last expression is
valid for all $x\in I$, then $limsup\vert\frac{
V^{(l+1)}(x_0)}{V^{(l)}(x_0)}\vert =0$. This condition is not so
stringent as it may seem, for instance, the potential of a
harmonic oscillator satisfies it, i.e., $V^{(3)}(x) =0,~~\forall
x\in\Re$.
\bigskip

\section{Validity regions of the semiclassical approximation}
\bigskip
\bigskip

Let us now proceed to define the family of potentials which will
be addressed in the present work.

Firstly, if $x\leq 0$, the potential becomes infinite. Secondly,
it is a monotonically increasing function and

\begin{equation}
limV(x)\rightarrow 0,~~as~~x\rightarrow 0^+\label{Potential1}
\end{equation}

Additionally, it possesses bound states, in such a way that the
corresponding energy eigenvalues are $\Bigl\{E_0, E_1,
....\Bigr\}$, where $E_l<E_n$, if $l<n$.

Let us now suppose that the first and second derivatives of the
potential do not vanish $\forall x\in\Re^+$. This condition is
introduced in order to have in our family the case of a truncated
harmonic oscillator. The case of a particle freely falling in a
homogeneous gravitational field will also be addressed here,
though for this potential the second derivative vanishes we may
analyze it. The possibility of resorting to the WKB method, using
a linear approximation for the potential energy at the classical
turning point $x_0$, is feasible at those points $x$ such that
they fulfill (\ref{Interval1}) and (\ref{WKB1}). These two
conditions imply

\begin{equation}
\alpha^2\vert\frac{ V''(x_0)}{2}\frac{x_0 - x}{\alpha}\vert
>
\Bigl[\sqrt{\frac{\hbar^2}{8m}}\alpha\frac{V''(x_0)}{2}\Bigr]^{2/3}\vert
1 - \frac{4}{3}\frac{x_0 - x}{\alpha}\vert. \label{Condition1}
\end{equation}

Here we have defined the length parameter

\begin{equation}
\alpha = 2\vert V'(x_0)/V''(x_0) \vert. \label{Distance1}
\end{equation}

This assertion may be rephrased stating that WKB can not be
applied at a point $x$ in the classical region (the one is defined
by the interval $J= (0, x_0]$) if the following inequality holds

\begin{equation}
\alpha^2\vert\frac{ V''(x_0)}{2}\frac{x_0 - x}{\alpha}\vert \leq
\Bigl[\sqrt{\frac{\hbar^2}{8m}}\alpha\frac{V''(x_0)}{2}\Bigr]^{2/3}\vert
1 - \frac{4}{3}\frac{x_0 - x}{\alpha}\vert. \label{Condition2}
\end{equation}

If the use of a linear approximation to the potential is
inconsistent we could try to employ the WKB formalism considering
higher--order derivatives in the approximate Schr\"odinger
equation, and we would end up with an inequality (though a more
complicated one) and a new length parameter, the one would be
given by the expression $\alpha = (l+1)\vert
V^{(l)}(x_0)/V^{(l+1)}(x_0)\vert$.

 Let us now analyze the implications of the breakdown of the linear approximation, and study
 its consequences, if any, upon the energy eigenvalues.

Defining

\begin{equation}
\gamma = \Bigl\{\frac{\hbar^2}{4m\alpha^4V''(x_0)}\Bigr\}^{1/3},
\label{Definition1}
\end{equation}

\begin{equation}
z= \frac{x_0 - x}{\alpha}. \label{Definition2}
\end{equation}

We may cast (\ref{Condition2}) as follows

\begin{equation}
\vert z\vert \leq \gamma\vert 1- 4z/3\vert. \label{Condition3}
\end{equation}

We consider now the classical region $J$, and restrict $x$ to it,
then $z\in [0, x_0/\alpha]$.

This last condition allows us to define the following two
functions

\begin{equation}
f: [0, x_0/\alpha]\rightarrow \Re, \label{Function1}
\end{equation}

\begin{equation}
g: [0, x_0/\alpha]\rightarrow \Re, \label{Function2}
\end{equation}

\begin{equation}
f(z) = z, \label{Function11}
\end{equation}

\begin{equation}
g(z) = \gamma\vert 1- 4z/3\vert. \label{Function21}
\end{equation}

It is readily seen that $\forall z\in[0, x_0/\alpha]$ we have that
$f(z)\geq 0$. On the other hand, our function $g(z)$ may change
sign in this interval. Indeed,

\begin{equation}
\gamma\vert 1- 4z/3\vert=
\left\{\begin{array}{cl} \gamma[1- 4z/3], & \mbox{when $z\leq3/4 $}\\
\gamma[-1+ 4z/3], &\mbox{when $z\geq 3/4 $}
\end{array}\right
.\label{Function22}
\end{equation}

We have two possibilities

\subsection{First Case ($x_0/\alpha \leq 3/4$)}
\bigskip
\bigskip

Then the condition that implies the breakdown of WKB reads

\begin{equation}
z\leq\gamma[1- 4z/3]. \label{Condition11}
\end{equation}

The border that divides the regions, in which WKB is valid from
the one in which it is not valid, is given by

\begin{equation}
z = \gamma[1- 4z/3], \label{Condition12}
\end{equation}

\begin{equation}
z = \frac{\gamma}{1+ 4\gamma/3}. \label{Condition13}
\end{equation}

In other words, in the region $z\in[0, \frac{\gamma}{1+
4\gamma/3}]$ WKB cannot be used, and in $z\in[\frac{\gamma}{1+
4\gamma/3}, x_0/\alpha]$ WKB is valid.

Notice that if the interval $z\in[\frac{\gamma}{1+ 4\gamma/3},
x_0/\alpha]$ becomes one point, then in the region $x\in(0, x_0]$
WKB cannot be used. This happens when

\begin{equation}
\frac{\gamma}{1+ 4\gamma/3} = x_0/\alpha, \label{Condition14}
\end{equation}

This condition can be cast as follows

\begin{equation}
\frac{\hbar^2}{8m} = \frac{\alpha
x^3_0}{2}V^{''}(x_0)\Bigl[1-\frac{4x_0}{3\alpha}\Bigr]^{-3}.
\label{Condition15}
\end{equation}

The roughest approximation allow us to rewrite (\ref{Condition15})
as

\begin{equation}
\frac{\hbar^2}{8m} = x^3_0V'(x_0). \label{Condition16}
\end{equation}

Additionally, we have the condition $x_0/\alpha \leq 3/4$, which
implies

\begin{equation}
x_0 \leq\frac{3V'(x_0)}{2V''(x_0)}. \label{Condition17}
\end{equation}

Consider now the following potential ($\beta>0$)

\begin{equation}
V(x)=
\left\{\begin{array}{cl} \beta x^n, & \mbox{when $x>0 $}\\
\infty, &\mbox{when $x\leq 0 $}
\end{array}\right.
.\label{Partpotential1}
\end{equation}

(\ref{Condition17}) implies that for this kind of potentials

\begin{equation}
n\leq 5/2. \label{Condition18}
\end{equation}

Notice that the fulfillment of (\ref{Condition17}) for the
aforementioned family of potentials does not involve the value of
$\beta$, only of $n$. Under this condition we may find some
important potentials of quantum physics, namely, the harmonic
oscillator ($n=2$) and a particle freely falling in a homogeneous
gravitational field ($n=1$).

The value of $x_0$, using (\ref{Condition16}) and
(\ref{Partpotential1}), is given by

\begin{equation}
x_0 = \Bigl(\frac{\hbar^2}{8\beta mn}\Bigr)^{1/(n+2)}.
\label{Condition19}
\end{equation}

From this last expression, additionally, we find the following
energy

\begin{equation}
E^{(1)}_{cl} = \beta x^{n}_0 = \frac{\hbar^2}{8mnx^2_0}.
\label{ClassicalEnergy2}
\end{equation}

The meaning of this energy is the following one. For this kind of
potential this is the minimum energy that a particle can have in
order to, with a linear approximation to the potential, be able to
employ WKB. Indeed, notice that $x_0$ is the smallest value of the
coordinate at which the aforementioned formalism can be employed,
and since we have assumed, from the very beginning, a
monotonically increasing potential, then we deduce that if a
particle has an energy smaller than (\ref{ClassicalEnergy2}), then
WKB cannot be employed.

At this point we may wonder if this energy has some physical
meaning. We will show that it is a lower bound to the
eigenenergies of the corresponding bound states, i.e., all the
eigenenergies stemming from the solution to the corresponding
Schr\"odinger equation, and related to bound states, are larger or
equal to $E^{(1)}_{cl}$. This assertion will be proved in section
IV.
\bigskip
\bigskip

\subsection{Second Case ($x_0/\alpha \geq 3/4$)}
\bigskip
\bigskip

In this case the breakdown of WKB involves expression
(\ref{Function22}), and clearly the first point of interest is
related to the fulfillment of expression (\ref{Condition12}). In
other words, once again, in the region $z\in[0, \frac{\gamma}{1+
4\gamma/3}]$ WKB cannot be used. But, since we have that
$x_0/\alpha \geq 3/4$, now the breakdown of WKB can include a new
interval, the one is absent when $x_0/\alpha \leq 3/4$. Indeed,
consider now the possibility in which

\begin{equation}
z =\gamma[-1+ 4z/3]. \label{Condition21}
\end{equation}

This happens when

\begin{equation}
z = \frac{\gamma}{-1 + 4\gamma/3}. \label{Condition22}
\end{equation}

Clearly, the case $4\gamma/3=1$ has to be discarded in this last
equation. Nevertheless, we must also analyze the case in which
$4\gamma/3=1$. Expression (\ref{Condition11}) tells us that in the
region $z\in[\frac{\gamma}{1+ 4\gamma/3}, x_0/\alpha]$ WKB can be
used, this is when $4\gamma/3=1$ This is easy to understand since
the slope of the two straight lines, $f(z) = z$ and $g(z) =
\gamma[-1+ 4z/3]$, is the same, $4\gamma/3=1$.

Harking back to the case $4\gamma/3\not=1$, we conclude that if
$\frac{\gamma}{-1 + 4\gamma/3}\leq x_0/\alpha$, then we have three
different regions, namely:

(i) If $z\in[0, \frac{\gamma}{1+ 4\gamma/3}]$ WKB cannot be used;
and it is related to the fact that we are too close to the
classical turning point and in consequence (\ref{WKB1}) is not
valid.

(ii) If $z\in[\frac{\gamma}{1+ 4\gamma/3}, \frac{\gamma}{4\gamma/3
-1}]$ WKB can be used;

(iii) Finally, when $z\in[\frac{\gamma}{4\gamma/3 -1},
x_0/\alpha]$ WKB loses, once again, its validity, but this time
the breakdown of the method emerges because the approximation
introduced for the Schr\"odinger equation at the classical turning
point is violated, see expression (\ref{Distance1}).

Of course, this last possibility occurs only when

\begin{equation}
3/4<\frac{\gamma}{4\gamma/3 -1}\leq x_0/\alpha.
\label{Condition23}
\end{equation}

This last condition can be cast as follows

\begin{equation}
\frac{\hbar^2}{8m}\geq\Bigl(3/4\Bigr)^3\alpha^4\frac{V''(x_0)}{2}.
\label{Condition24}
\end{equation}

Consider now the case in which the potential is given by
(\ref{Partpotential1}). One of our previous results, see
(\ref{Condition18}), imposes as condition $n>5/2$. Then we obtain
as condition for the existence of the region defined by
$z\in[\frac{\gamma}{4\gamma/3 -1}, x_0/\alpha]$ the following
inequality

\begin{equation}
\frac{\hbar^2}{8mx^2_0}\geq\Bigl(\frac{2}{3[n-1]}\Bigr)^3nV(x_0).
\label{Condition25}
\end{equation}
\bigskip
\bigskip

This last expression allows us to define the following energy

\begin{equation}
E^{(1)}_{cl} =
\Bigl(\frac{3[n-1]}{2}\Bigr)^3\frac{\hbar^2}{8mnx^2_0}.
\label{Condition26}
\end{equation}
\bigskip
\bigskip

\subsection{Comparison between the two cases}
\bigskip
\bigskip

If $x_0/\alpha \geq 3/4$, then we cannot obtain the value of $n$
without knowing also the value of $\beta$ (the one appears in
$V(x_0)$), as expression (\ref{Condition25}) clearly shows. This
does not happen in the first situation ($x_0/\alpha \leq 3/4$),
see expression (\ref{Condition18}), in which the condition upon
$n$ is independent from the value of $\beta$, though $x_0$ does
involve $\beta$, see (\ref{Condition19}).

An additional difference between these two cases lies in the fact
that if ($x_0/\alpha \leq 3/4$) then we may find a family of
potentials such that the analysis contains only one region, see
expression (\ref{Condition15}), in which WKB is not valid (in the
interval $(0, x_0)$ the conditions imposed by WKB cannot be
satisfied), whereas in the other situation the analysis shows the
possibility of having two or three regions, and in one of them WKB
can always be employed.

As a byproduct we have obtained for the first case an energy, see
expression (\ref{ClassicalEnergy2}) such that for energies smaller
than it the WKB method cannot be used. Similarly, for the second
situation, expression (\ref{Condition26}) defines an energy such
that if the energy of our particle is smaller than that provided
by (\ref{Condition26}), then WKB cannot be used. In other words,
in both cases we find a lower bound for the validity of the
method, the one includes a linear approximation for the
Schro\"odinger equation at the classical turning point.
\bigskip
\bigskip

\section{Lower bound for the ground state energy}
\bigskip
\bigskip

Let us now consider the classical energy $E^{(1)}_{cl} =V(x_0)$
obtained previously, see expressions (\ref{ClassicalEnergy2}) and
(\ref{Condition26}). The meaning of this energy is the following
one. For the kind of potential under consideration this is the
minimum energy that a particle can have in order to, with a linear
approximation to the potential, be able to employ WKB. Indeed,
notice that $x_0$ is the smallest value of the coordinate at which
the aforementioned formalism can be employed, and since we have
assumed, from the very beginning, a monotonically increasing
potential, then we deduce that if a particle has an energy smaller
than (\ref{ClassicalEnergy2}) or (\ref{Condition26}), then this
energy can not be obtained via WKB.

At this point we may wonder if this energy has some physical
meaning. Let us now answer this question. We will conjecture that
if

\begin{equation}
E^{(1)}_{cl}\leq \frac{\hbar^2}{8mx^2_0}, \label{Condition27}
\end{equation}

\noindent then it is a lower bound to the eigenenergies of the
corresponding bound states, i.e., all the eigenenergies stemming
from the solution to the corresponding Schr\"odinger equation, and
related to bound states, are larger or equal to $E^{(1)}_{cl}$.

Let us now consider the lower energy eigenstates of a quantum
system subject to a potential satisfying our requirements. Under
this condition WKB may be used, though it provides energies that
are not a good approximation to the correct ones. We now proceed
to prove that when $E^{(1)}_{cl}$ satisfies (\ref{Condition27}),
then it does define a lower bound to the ground energy of the
system. Consider the energy $E_{0}$ of the ground state and assume
that it intersects the potential at $x_g$, i.e., $E_0 = V(x_g)$.
Here we have two possibilities:

(i) $x_g$ lies on the right--hand side of $x_0$. In this case
$E_0$ can be obtained via WKB, and since $E^{(1)}_{cl}$ is the
smallest energy that allows the use of this approach, then
$E^{(1)}_{cl}\leq E_0$, and we are done;

(ii) Consider the remaining possibility, $x_g <x_0$. Since we have
assumed a monotonically increasing potential, then this last
condition implies that $V(x_g) < V(x_0)$.

The idea here is to proceed by contradiction, namely, assume that
$x_g <x_0$, a fact that implies $V(x_g) < V(x_0)$, and resorting
to the fundamentals of quantum mechanics we will show that this
assumption leads to an inconsistency for those potentials
fulfilling (\ref{Condition27}).

According to quantum mechanics the distance that the wave function
(of a bound state) tunnels inside the classically forbidden region
becomes larger as the energy grows. Indeed, in the aforementioned
region the wavefunction behaves as
$\psi(x)\approx\exp\Bigl\{-\frac{1}{\hbar}\int_a^{x}\sqrt{2m[V(z)
- E]}dz\Bigr\}$ \cite{[4]}.

Let us now recall the uncertainty relation

\begin{equation}
\Delta x\Delta p \geq \frac{\hbar}{2}. \label{Uncertainty1}
\end{equation}

For the ground state the tunnelling distance is the smallest one,
and if $x_g$ denotes the intersection of the ground energy with
potential, then $(0, x_g]$ is the classical region, and we may
state that

\begin{equation}
\Delta x\sim x_g. \label{Delta1}
\end{equation}

This last expression says that the probability of finding the
particle in the classically forbidden region is almost zero, and
in consequence, the uncertainty region associated to the position
of the particle is related to the classical region, in which the
wavefunction has an oscillatory behavior. Bearing this in mind the
uncertainty relation can be cast in the following form

\begin{equation}
x_g\sqrt{2m[E_0 - <V>] - (<p>)^2)}\geq \frac{\hbar}{2}.
\label{Uncertainty2}
\end{equation}

Where the averages are calculated with the wavefunction of the
ground state. Since we have assumed that $V(x)\geq 0,~~\forall
x\in\Re^+$, then (\ref{Uncertainty2}) implies

\begin{equation}
x_g\sqrt{2mE_0}\geq \frac{\hbar}{2}. \label{Uncertainty3}
\end{equation}

Then the ground state energy satisfies the condition

\begin{equation}
E_0 \geq \frac{\hbar^2}{8mx^2_g}. \label{Ground1}
\end{equation}

And since our initial hypothesis has been the condition $x_g<x_0$,
then

\begin{equation}
\frac{\hbar^2}{8mx^2_0}\leq \frac{\hbar^2}{8mx^2_g}.
\label{Uncertainty4}
\end{equation}

Nevertheless, we have assumed from square one the validity of
(\ref{Condition27}), and hence we find a contradiction. Indeed,
our three initial conditions (expression (\ref{Condition27}) and
$x_g<x_0$ and the uncertainty relation) imply $E^{(1)}_{cl}\leq
E_0$. Nevertheless, since the potential has been assumed
monotonically increasing, then $x_g<x_0\Rightarrow V(x_g) <
V(x_0)$, and in consequence, we find that the simultaneous use of
these three conditions is, logically, inconsistent, and therefore
the case $x_g<x_0$ has to be discarded, if expression
(\ref{Condition27}) and the uncertainty relation are to be kept.
Clearly, we conclude that $E^{(1)}_{cl}$ provides a lower bound
for the ground energy.
 We now proceed to analyze if any of our two possible cases satisfies (\ref{Condition27}).

\bigskip
\subsection{First case}
\bigskip
\bigskip

Consider now a potential given by (\ref{Partpotential1}), with
$1\leq n\leq 5/2$. Then, $E^{(1)}_{cl} =\frac{\hbar^2}{8mnx^2_0}$,
and condition (\ref{Condition27}) is fulfilled.

In other words, if $1\leq n\leq5/2$, then $E^{(1)}_{cl}$ does
provide a lower bound to the ground energy of the corresponding
system.

\bigskip
\subsection{Second case}
\bigskip
\bigskip

The remaining possibility, $n>5/2$, cannot be analyzed in the
context of our approach. Indeed, notice that now we require the
fulfillment of the following condition

\begin{equation}
\frac{\hbar^2}{8mx^2_0}\geq\Bigl(\frac{3[n-1]}{2}\Bigr)^3\frac{\hbar^2}{8mnx^2_0}.
\label{Ground3}
\end{equation}

This last condition can be cast as

\begin{equation} \Bigl(\frac{3}{2}\Bigr)^3\leq\frac{n}{(n-1)^3}.
\label{Ground4}
\end{equation}

Clearly, not only (\ref{Ground4}) has to be satisfied,
additionally $n>5/2$. A fleeting glimpse shows that this is
impossible, in other words, we cannot use the present approach to
find a lower bound for potential of the form $V(x) = \beta x^n$,
if $n>5/2$.
\bigskip
\bigskip

\section{Some particular potentials}
\bigskip
\bigskip

Notice that under the condition $n\leq 5/2$ we may find the
potential associated to a harmonic oscillator, or a particle
freely falling in a homogeneous gravitational field. We know
proceed to calculate the bounds for these two potentials that our
method provides. We use these two cases to corroborated our
predictions, since we need potentials whose exact eigenenergies
are already known.

\subsection{Harmonic Oscillator}
\bigskip

A particular case of the situation just mentioned is the potential
associated to a harmonic oscillator. Notice that for this case the
condition $n<5/2$ is fulfilled, and in consequence, if our
argument is valid, then we expect to obtain a lower bound to the
ground energy.

\begin{equation}
V(x)=
\left\{\begin{array}{cl} \frac{m}{2}\omega^2x^2, & \mbox{when $x>0 $}\\
\infty, &\mbox{when $x<0 $}
\end{array}\right.
.\label{Relation3}
\end{equation}

In this case (see expression (\ref{Condition19}))

\begin{equation}x_0 = (\frac{1}{8})^{1/4}\lambda_0,\label{Harmonic1}
\end{equation}

\begin{equation}
\lambda_0 = \sqrt{\frac{\hbar}{m\omega}} ,\label{Harmonic2}
\end{equation}

\begin{equation}
 E^{(1)}_{cl} = \frac{1}{2\sqrt{8}}\hbar\omega\label{Harmonic3}
\end{equation}

The eigenenergies of a harmonic oscillator read $E_n =
\hbar\omega[(2n+1) +1/2]$, in this case $n$ is an integer.
Clearly, since we have the truncated harmonic oscillator
potential, only the odd wave functions are allowed and, in
consequence, the ground state is given by $E_0 =
\frac{3}{2}\hbar\omega$. Here we have that, indeed,
$E^{(1)}_{cl}<E_0$.
\bigskip

\subsection{Free Fall}
\bigskip

Another interesting situation in quantum mechanics, the one has
exact solution, is the case of a particle falling in a homogeneous
gravitational field \cite{[5]}. Here the potential satisfies the
condition

\begin{equation}
V(x)=
\left\{\begin{array}{cl} mgx, & \mbox{when $x>0 $}\\
\infty, &\mbox{when $x\leq 0 $}
\end{array}\right.
.\label{Freefall1}
\end{equation}

 The eigenenergies are given by ($n\in\mathbb{N}$)

\begin{equation}
E_n=
\frac{\hbar^2}{2ml^2}\Bigl\{\frac{3\pi}{4}\Bigl[2n-\frac{1}{2}\Bigr]\Bigl\}^{2/3},\label{Freefall2}
\end{equation}

\begin{equation}
\frac{1}{l^3}= \frac{2m^2g}{\hbar^2}.\label{Freefall3}
\end{equation}

For this type of potential the second derivative vanishes in the
region $(0, \infty)$, and hence, the condition that determines the
minimum distance $x_0$ is given by

\begin{equation}
x_0
\leq\frac{1}{2}\Bigl(\frac{\hbar^2}{m^2g}\Bigr)^{1/3}.\label{Freefall4}
\end{equation}

Since the potential is linear, then, clearly, the linear
approximation to Schr\"odinger equation is always valid, and the
breakdown of the method is only related to the violation of
(\ref{WKB1}).

The ensuing energy reads

\begin{equation}
E^{(1)}_{cl}= \frac{\hbar^2}{\sqrt[3]{32}\,ml^2}.\label{Freefall5}
\end{equation}

Notice that the energy eigenvalues can be cast as follows

\begin{equation}
E_n=
E^{(1)}_{cl}\Bigl\{\frac{3\pi}{2}\Bigl[2n-\frac{1}{2}\Bigr]\Bigl\}^{2/3}.\label{Freefall6}
\end{equation}

The ground state corresponds to $n=0$, and then (\ref{Freefall6})
becomes

\begin{equation}
E_0=
E^{(1)}_{cl}\Bigl\{-\frac{3\pi}{4}\Bigl\}^{2/3}.\label{Freefall7}
\end{equation}

Clearly, once again, the condition, $E_0>E^{(1)}_{cl}$ is fulfilled,
$\Bigl\{-\frac{3\pi}{4}\Bigl\}^{2/3}>1$.

\bigskip

\section{Conclusions}
\bigskip
\bigskip

In the present work it has been shown that the so--called WKB
approximation method involves two length parameters which have to
be introduced in the calculation of the relevant variables.
Usually, only one of these length parameters is considered in the
text books on quantum mechanics \cite{[1], [2]}, the second one is
always neglected. These parameters are related to different
conditions that have to be fulfilled in order to have a
mathematically consistent approximation method: (i) the first one
tells us that we cannot be very close to the classical turning
point, otherwise the approximate wavefunction diverges \cite{[1],
[2]}, see expression (\ref{WKB1}); (ii) whereas the second length
parameter appears when we recognize that the comparison between
the solution to the approximate Schr\"odinger equation and the WKB
wavefunction has to be carried out within the validity region of
the aforementioned motion equation, see expression
(\ref{Interval1}).

Both conditions have been taken into account and it has been shown
that a careful analysis of them leads us to some interesting
facts. For instance, it is possible to find a lower bound to the
ground energy for some potentials, i.e., those whose behavior goes
like $V(x) = \beta x^n$, $\beta >0$ and $1\leq n\leq 5/2$. Joining
the present conclusions with the so--called Rayleigh--Ritz method
allows us to find an interval in which the ground energy of a
one--dimensional system has to lie. Additionally, some examples of
the approach have been explicitly calculated, for instance, the
case of a harmonic oscillator, or a particle freely falling in a
homogeneous gravitational field, and it has been shown that the
energy deduced with our arguments is, indeed, smaller than the
ground energy of the corresponding system.

\begin{acknowledgments}
 A. C. was supported by CONACYT Grant 47000--F, whereas L. B. by CONACYT Scholarship number
 188230. A. C. would like to thank A.A. Cuevas--Sosa for useful discussions
and literature hints.
\end{acknowledgments}

\end{document}